\documentclass[a4paper,twoside]{article}

\usepackage{epsfig}
\usepackage{subcaption}
\usepackage{calc}
\usepackage{amssymb}
\usepackage{amstext}
\usepackage{amsmath}
\usepackage{amsthm}
\usepackage{multicol}
\usepackage{pslatex}
\usepackage{apalike}
\usepackage[bottom]{footmisc}

\usepackage{xcolor} 

\usepackage{hyperref}
\usepackage{booktabs}
\usepackage{float}

\usepackage{SCITEPRESS}

\begin{document}

\title{Combining Generators of Adversarial Malware Examples to Increase Evasion Rate}

\author{\authorname{
Matouš Kozák\sup{1}\orcidAuthor{0000-0001-8329-7572}, 
Martin Jureček\sup{1}\orcidAuthor{0000-0002-6546-8953}}
\affiliation{\sup{1}Faculty of Information Technology, Czech Technical University in Prague, Thákurova 9, Prague, Czechia}
\email{\{kozakmat, martin.jurecek\}@fit.cvut.cz}
}


\abstract{Antivirus developers are increasingly embracing machine learning as a key component of malware defense. While machine learning achieves cutting-edge outcomes in many fields, it also has weaknesses that are exploited by several adversarial attack techniques. Many authors have presented both white-box and black-box generators of adversarial malware examples capable of bypassing malware detectors with varying success. We propose to combine contemporary generators in order to increase their potential. Combining different generators can create more sophisticated adversarial examples that are more likely to evade anti-malware tools. We demonstrated this technique on five well-known generators and recorded promising results. The best-performing combination of AMG-random and MAB-Malware generators achieved an average evasion rate of 15.9\% against top-tier antivirus products. This represents an average improvement of more than 36\% and 627\% over using only the AMG-random and MAB-Malware generators, respectively. The generator that benefited the most from having another generator follow its procedure was the FGSM injection attack, which improved the evasion rate on average between 91.97\% and 1,304.73\%, depending on the second generator used. These results demonstrate that combining different generators can significantly improve their effectiveness against leading antivirus programs.}

\onecolumn \maketitle \normalsize \setcounter{footnote}{0} \vfill

\section{\uppercase{Introduction}} \label{sec:introduction}
Malware is software performing malicious actions on infected computers. As more and more of our lives become digital, protecting our devices becomes increasingly important. Cybersecurity professionals are striving to improve the detection capabilities of their antivirus (AV) products by inventing new defense mechanisms \cite{gibert2020rise}. Nonetheless, their rivals are progressing at a comparable, if not faster, pace, making malware detection a never-ending fight.

Leading AV programs use both static and dynamic analysis. Static analysis techniques usually rely on byte sequences (signatures) kept in a database. Signatures accurately and quickly detect known harmful files, but their fundamental drawback is their incapability to classify zero-day or obfuscated malware. Even minor changes to malware files may cause the signature to change, thus rendering them undetectable by static analysis. In contrast, dynamic analysis methods include behavior-based algorithms that search for patterns of behavior that can be used to discover unknown and obfuscated malware samples, albeit at a higher cost of executing malware in a safe environment \cite{aslan2020comprehensive}.


While traditional signature-based static analysis cannot detect zero-day malware, incorporating machine learning (ML)-based malware detectors gives encouraging results \cite{comar2013combining}. However, ML models are vulnerable to adversarial examples (AEs), e.g., slightly changing a malicious file can cause its feature vector to mimic some of the benign files' feature vectors \cite{papernot2016limitations}. As a result, malware detectors may make inaccurate predictions.

We propose a novel adversarial attack strategy that combines generators of adversarial malware examples. By combining different generators, we aim to create more sophisticated AEs capable of bypassing top-tier AV products. Our method works at the level of samples, i.e., the output of a combination of generators is a functional malware binary.

For various reasons, we focus our work on attacking static malware analysis. To begin, dynamic analysis requires running malware inside a secure environment and documenting its behavior, which is both time-consuming and technically challenging. Next, to our best knowledge, there is no successful implementation of AE generators targeting dynamic malware analysis, which we could use in our method. Furthermore, malware authors can use sandbox evasion techniques such as detecting that their malware is running in a controlled environment and ceasing its dangerous behavior \cite{erko2022malware,yuceel2022}. Finally, static detection is typically the initial line of defense against malicious threats, making it an essential component of any anti-malware tool.

\paragraph*{The outline of the paper}
\begin{itemize}
	\item In Section \ref{sec:background}, we establish the necessary background by briefly introducing adversarial machine learning and portable executable file format.
	\item In Section \ref{sec:related_work}, we summarize related work in the field of generating adversarial malware examples.
	\item In Section \ref{sec:proposed_method}, we define our method in detail. From a step-by-step description of combining adversarial malware generators to a comprehensive overview of each AE generator we used.
	\item In Section \ref{sec:evaluation}, we introduce our experiment's setup, dataset, and routine used. Finally, we demonstrate the achieved results.
	\item In Section \ref{sec:conclusion}, we summarize our contributions and make recommendations for future research.
\end{itemize}

\section{\uppercase{Background}} \label{sec:background}
In this section, we briefly introduce the key concepts for understanding this paper by describing adversarial machine learning with respect to malware detection and Portable Executable file format used on Windows operating systems.

\subsection{Adversarial Machine Learning} \label{subsec:adversarial_ML}
In recent years, we have seen an increase in the popularity of machine learning algorithms in a variety of domains, such as advertisement recommendation, image classification, and Go playing, where ML models achieve cutting-edge results \cite{marius2021sota,silver2017mastering}. However, in other areas, such as self-driving cars or disease diagnosis, both the general public and researchers remain skeptical of these models' decisions \cite{juravle2020trust,edmonds2020three}. One reason for skepticism about ML models is the unexplainable nature of their decisions and the following possible fragility and bias of the ML model \cite{leilani2018explaining}. As a result, ML systems can be vulnerable to minor changes exploited by adversarial attacks \cite{goodfellow2015explaining,papernot2016limitations}.

\textit{Adversarial machine learning} is a branch of machine learning that focuses on enhancing ML systems' resistance against adversarial attacks both from the outside (evasion attacks) and from the inside (data poisoning). An adversarial attack is a well-planned activity designed to deceive the ML model. The victim model is also known as a target model, and the attacker is referred to as an adversary. Nonetheless, in contemporary literature, the terms attacker and adversary are used interchangeably. The input responsible for fooling the target model is called an \textit{adversarial example (AE)}.

In the malware detection domain, adversarial machine learning is usually used to force anti-malware tools to misclassify malware as benign. An arms race between adversaries and antivirus defenders has evolved as a result of the expansion of machine learning employed in malware detection \cite{ucci2019survey}. Malware adversarial learning focuses on how malware can trick malware detection models and how to develop malware detectors resistant to AEs.

An adversarial attack's success is limited by the amount of information accessible about the target system \cite{huang2011adversarial}. A \textit{white-box} scenario occurs when the attacker gets access to the target system and may study its internal configuration or training datasets. A \textit{black-box} scenario, on the other hand, happens when the adversary has minimal knowledge about the victim detector, usually only in the form of the detector's final prediction, e.g., malware/benign label for each submitted sample. In between these two extremes is a \textit{grey-box} scenario, in which the attacker has greater access to the system than in the black-box scenario, however, only to certain aspects of it. For instance, the attacker can access the model's feature space but not its training dataset. In the field of adversarial malware generation, the black-box scenario is the most practical, as the exact structure of the AV is usually unknown to the adversary.

\subsection{Portable Executable File Format}
\textit{Portable Executable (PE) file format} is widely used on Windows operating systems. This format is utilized by executable files (EXEs) or dynamically linked libraries (DLLs) on 32-bit and 64-bit systems and is structured as follows. The program begins with the MS-DOS header and stub program, which are now mostly obsolete and are only included for backward compatibility. The \verb|e_magic| (identifying the file as a MS-DOS executable) and \verb|e_lfanew| (the file offset of the Common Object File Format (COFF) file header) fields are the exception. Following is the \verb|signature| and the COFF file header, which contains information such as the target machine and section table size. The COFF file header is closely followed by the optional header, which includes, among other things, the necessary data directories. A section table with corresponding section data completes the program \cite{microsoftPE}.

\section{\uppercase{Related Work}} \label{sec:related_work}
This section summarizes related publications that address the development of adversarial malware attacks. We begin by describing the works that leverage \textit{gradient-based} techniques to exploit the back-propagation algorithm, a widely used algorithm in training deep neural networks, by computing the necessary perturbations to mislead the target classifier \cite{goodfellow2015explaining,papernot2016limitations}. Next, we show studies employing \textit{reinforcement learning-based} attacks where reinforcement learning agent equipped with a set of actions in the form of binary file manipulations attempts to find a sequence of modifications leading to misclassification \cite{anderson2018learning}. Finally, we highlight a few publications related to adversarial malware attacks that do not fall into either of the two categories above.


\subsection{Gradient-Based Attacks}
In \cite{grosse2017adversarial}, a gradient-based attack against a self-made Android malware detector was proposed. The necessary perturbation for the feature vector containing features extracted from the Android manifest file was calculated using the gradient descent algorithm. The authors recorded an evasion rate of up to 63\% against their deep neural network classifier.

Many authors proposed gradient-based attacks against the MalConv malware classifier \cite{raff2017malware}. For example, in \cite{kolosnjaji2018adversarial}, the authors perturbed the file's overlay and achieved a 60\% evasion rate while altering less than 1\% of total bytes. 

Further, the authors of \cite{kreuk2018deceiving} introduced the injection of small chunks of bytes (payload) into unused regions or at the end of the PE file, achieving a 99\% evasion rate against MalConv while limiting the payload size to less than 1,000 bytes. 

Explanation techniques used on the MalConv detector and subsequent attack perturbing obsolete parts of the MS-DOS header were introduced in \cite{demetrio2019explaining}, reaching an evasion rate of over 86\%.

A complex approach that treats EXE binaries as images was introduced in \cite{yang2021deepmal}. The authors computed the necessary perturbations in the embedding space of a convolutional neural network-based malware detector and later mapped them to corresponding sections of the original input binary. Their attack decreased the accuracy of selected ML models by up to 94\%.

\subsection{Reinforcement Learning-Based Attacks}
While gradient-based attacks were originally intended for the image domain, the use of reinforcement learning agents is novel for the field of malicious AEs. This approach was pioneered by Anderson et al. in \cite{anderson2018learning}, introducing the actor-critic model capable of modifying raw binary files. The authors recorded an evasion rate of 24\% against the gradient-boosted decision tree (GBDT) detector trained on the EMBER dataset \cite{anderson2018EMBER}.

Fang et al. proposed two models in \cite{fang2020deepdetectnet}, a detector called DeepDetectNet and an AE generator called RLAttackNet. In a black-box setting, their generator, based on the deep q-network (DQN) algorithm, successfully evaded their classifier in 19.13\% of cases.

Song et al., the authors of the MAB-malware framework, employed a multi-armed bandit (MAB) agent while targeting commercial AVs, GBDT, and MalConv detectors \cite{song2022mab}. They demonstrated bypassing detection from commercial AVs with a high evasion rate of up to 48.3\%. The GBDT and MalConv detectors were successfully misled 74.4\%  and 97.7\% of times, respectively.

In \cite{kozak2022generation}, the authors used the DQN agent to generate adversarial malware examples. While targeting the GBDT and MalConv classifiers, their adversarial malware generator recorded an evasion rate of 68.64\% and 13.32\%, respectively. Further, they were the first to introduce a reverse scenario of generating adversarial benign examples and misled the GBDT and MalConv detectors in 3.45\% and 14.29\% of cases, respectively. While creating malicious AEs is more prevalent in current research, the increase in false positives of AV products would render them ineffective.

\begin{figure*}[ht!]
	\centering	
	\includegraphics[]{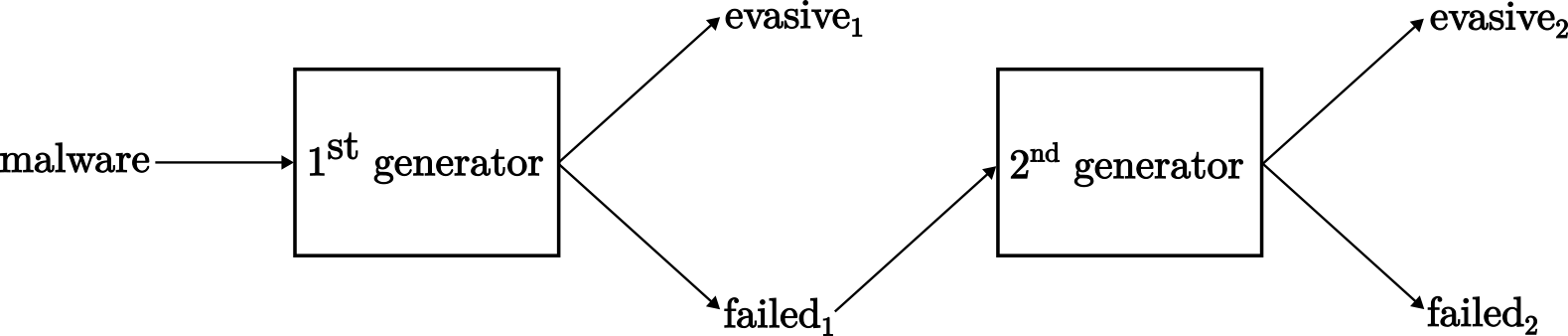} 
	\caption{Overview of our proposed method for generating adversarial examples via combining two generators.}
	\label{fig:method_diagram}
\end{figure*} 

\subsection{Other Methods}
To generate AEs, the authors of \cite{hu2017generating} presented MalGan, a generative adversarial network (GAN). They used a deep neural network substitute detector during training, i.e., the final target model was not used during training. The authors' results show strong attack transferability across the substitute and target models by recording a near-perfect evasion rate. Their work, however, is situated only in a feature space of extracted API calls, with no technique provided for transforming the adversarial feature vectors back into real-world EXEs.

Ebrahimi et al. suggested a recurrent neural network-based generative sequence-to-sequence language model. The model generates benign bytes, which are later appended to the end of malicious PE binaries \cite{ebrahimi2020binary}. Their proposed attack successfully evaded detection by the MalConv classifier on average in 73.24\% of cases across different malware families.

A genetic algorithm for generating adversarial malware examples was presented in \cite{demetrio2021functionality}. The AEs are constrained to maximize evasion against the GBDT classifier while minimizing the resulting file size. The authors achieved an evasion rate of up to 57\%.

\section{\uppercase{Proposed Method}} \label{sec:proposed_method}
In this section, we introduce a novel approach to improve current adversarial malware generators by combining them. Our method combines two separate AE generators and reuses non-evasive adversarial examples from the first generator as input to the second generator with the goal of crafting sophisticated malicious AEs capable of evading malware detectors. 

The overview of our proposed method is depicted in Figure \ref{fig:method_diagram}. Firstly, the first generator processes genuine malware samples, creating $evasive_1$ and $failed_1$ (i.e., non-evasive) sets. The $evasive_1$ examples are left untouched as they have successfully bypassed the given detector. The $failed_1$ samples, on the other hand, are passed as input to the second generator, and two new sets, $evasive_2$ and $failed_2$, are produced. The $failed_2$ samples are the resulting non-evasive samples that were unable to escape detection and the $evasive_2$ together with $evasive_1$ form the set of successful AEs crafted by combining two generators of adversarial malware examples. The division of AEs into $evasive$ and $failed$ groups is done by an independent classifier that is not part of any of the generators. In other words, when we attack a specific AV product (playing the role of independent classifier), we first send all generated AEs from the first generator to the AV detector, thus obtaining the $evasive_1$ and $failed_1$ sets. Next, we process the $failed_1$ files with the second generator, and the resulting AEs are resubmitted to the AV detector, resulting in $evasive_2$ and $failed_2$ sets. Note that the order of generators is essential, i.e., combining generators does not possess commutative property.

We studied combinations of five distinct adversarial generators which emit AEs targeted against MalConv \cite{raff2017malware} and EMBER GBDT \cite{anderson2018EMBER} classifiers. The EMBER GBDT model is a gradient-boosted decision tree that classifies inputs based on 2,381 extracted features from binary files. Binary files are parsed using the LIEF\footnote{\url{https://lief-project.github.io/}} library, and resulting extracted features include records such as information from PE headers, imported functions, section characteristics, byte histograms, and more. In contrast, MalConv is a convolution network that consumes directly raw bytes from binary executables truncated to 2,000,000 bytes (2 MB). Both models with pre-trained configurations are freely available on GitHub\footnote{\url{https://github.com/endgameinc/malware_evasion_competition}}.

\subsection{Generators of Adversarial Malware Examples}
For our work, we selected these five generators: MAB-Malware\footnote{\url{https://github.com/bitsecurerlab/MAB-malware}}, AMG\footnote{\url{https://github.com/matouskozak/AMG}} (trained and random versions), FGSM\footnote{\url{https://github.com/pralab/secml_malware}\label{footnote:secml}} and Partial DOS\footref{footnote:secml}. The first three generators work in pure black-box settings, whereas the latter two operate in a white-box manner. Despite the fact that we used white-box generators of adversarial malware, our proposed method is essentially a black-box attack because we do not focus on any of the target classifiers used by individual AE generators but on an independent classifier that is not part of our combined model.

\subsubsection{MAB-Malware}
\textit{MAB-Malware} is a generator of adversarial malicious examples utilizing a reinforcement learning algorithm called multi-armed bandit \cite{song2022mab}. In contrast with other reinforcement learning-based adversarial attacks, this algorithm works statelessly, meaning that the order of manipulations applied to genuine files is not considered. As a result, the algorithm operates in only two states, non-evasive and evasive, i.e., failure and success. The process of creating AEs consists of two main phases. Firstly, file modifications are applied until the target detector (MalConv) classifies the sample as benign or a count of 10 changes is reached. Secondly, the action minimization procedure removes unnecessary modifications, provided that the example remains evasive. If, on the other hand, the resulting example is not evasive, the application of modifications can be repeated up to 60 times.

\subsubsection{AMG}
\textit{Adversarial malware generator (AMG)} is a reinforcement learning-based generator for creating AEs \cite{kozak2023application}. This generator can operate in two settings. In the first variant, the generator uses the proximal policy optimization (PPO) algorithm to choose optimal actions based on the policy learned during training. In the second case, a random agent is deployed, i.e., no previous training is needed, and available actions are chosen at random. The possible actions are in the form of a predefined set of PE file manipulations that the agents repeatedly use until the evasion by the target classifier (EMBER GBDT) is accomplished or a maximum number of modifications, 50, is performed.

\subsubsection{FGSM}
\textit{Fast gradient sign method (FGSM)} is a gradient-based method for generating AEs introduced by Goodfellow et al. \cite{goodfellow2015explaining}. A modified version for the domain of malware samples is used where only a small chunk of bytes (payload) is perturbed and later inserted or appended to the original malware file \cite{kreuk2018deceiving}. At first, the payload is perturbed in the embedding space of the target classifier and later mapped back to the original binary. The perturbation of the payload is repeated until it escapes the detection of MalConv target classifier and is limited to a maximum of 100 times.

\subsubsection{Partial DOS}
\textit{Partial DOS} is another gradient-based algorithm capable of creating adversarial malware examples. This attack works by perturbing only bytes found in the MS-DOS header except for the \verb|e_magic| and the \verb|e_lfanew| fields \cite{demetrio2019explaining}. Same as with the FGSM algorithm, the perturbation is iterated until the AE bypasses the MalConv classifier and is limited to a maximum number of 100 rounds.

\section{\uppercase{Evaluation}} \label{sec:evaluation}
This section describes our setup used for experiments and how we evaluated our proposed combination of adversarial malware generators.

\subsection{Setup}
\textbf{Dataset}: In this paper, we used a dataset of PE malware binaries from the VirusShare\footnote{\url{https://virusshare.com/}} online repository, which we thank for access. Specifically, we used 2,000 samples from the VirusShare\_00454 dataset published on 01/02/2023. \\

\noindent \textbf{Computer setup}: Experiments were carried out on the NVIDIA DGX Station A100 server equipped with a single AMD 7742 processor with 64 cores, 512 GB of DDR4 system memory, and four NVIDIA A100 graphic cards with 40 GB of GPU memory.

\subsection{Experiments}
To evaluate our hypothesis that combining adversarial malware generators can significantly increase the probability of a successful adversarial attack, we first need to assess individual generators on some malware detectors. We picked the 10 best-rated antivirus programs as ranked by AV-Comparatives in their annual Summary Report from 2022 \cite{avcomparatives2022summary}. Note that we only list nine antivirus products in the following results because two selected AVs from the same company returned identical results. Further, we anonymize the names of chosen AVs to minimize possible misuse of this work.

The main metric used to evaluate malicious AEs is an evasion rate. This metric represents the ratio of adversarial malware examples incorrectly classified as benign to the total number of files tested and is computed as follows:

\begin{equation}
	evasion\ rate = \dfrac{misclassified}{total} \cdot 100\%
\end{equation}

\noindent where $total$ is the total amount of files submitted to the malware detector after discarding harmful files that were already mispredicted in their genuine form, i.e., before adversarial modification. 

\subsubsection{Baseline}
Firstly, we used the 2,000 malware samples mentioned earlier to generate AEs from all five tested generators of adversarial malware: MAB-Malware, AMG (PPO and random agents), FGSM, and Partial DOS. For generating AEs, we used the default configurations of each generator as specified by the respective authors. Next, we tested these AEs against publicly available versions of the above-mentioned top antivirus programs hosted on VirusTotal\footnote{\url{https://www.virustotal.com/}} website. Note that the AEs were generated against the corresponding target classifiers as described in Section \ref{sec:proposed_method} and not against these AV programs.

\begin{table*}[ht!]
	\centering
	\caption{Evasion rates of individual AE generators against selected AVs.}
	\label{table:individual_AE_generators}
	\begin{tabular}{@{}l|ccccccccc@{}}
\toprule
            & AV-1 & AV-2  & AV-3  & AV-4 & AV-5 & AV-6  & AV-7 & AV-8 & AV-9  \\ \midrule
MAB-Malware & 1.18 & 1.16  & 2.9   & 1.29 & 1.21 & 7.1   & 0.96 & 2.82 & 1.06  \\
AMG-PPO     & 0.41 & 2.75  & 2.41  & 2.84 & 1.79 & 3.96  & 2.39 & 1.9  & 3.69  \\
AMG-random  & \textbf{2.15} & \textbf{11.74} & \textbf{12.88} & \textbf{9.86} & \textbf{9.22} & \textbf{34.48} & \textbf{9.37} & \textbf{3.54} & \textbf{11.87} \\
FGSM        & 0.26 & 0.16  & 1.23  & 0.32 & 0.21 & 1.93  & 0.21 & 1.69 & 0.11  \\
Partial-DOS & 0.36 & 1.43  & 0.64  & 6.6  & 0.95 & 7.91  & 0.59 & 2.67 & 1.64  \\ \midrule 
average         & 0.87  & 3.45        & 4.01      & 4.18  & 2.68  & 11.08  & 2.7        & 2.52     & 3.67  \\ \bottomrule

	\end{tabular}
\end{table*}


We present the baseline results in Table \ref{table:individual_AE_generators}, which contains evasion rates for all AVs. Surprisingly, the random AMG agent outperformed other specialized AE generators, even its trained compatriot AMG with the PPO algorithm. The highest evasion rate of 34.48\% was recorded by the random AMG algorithm against the AV-6 detector, which performed the worst among the tested AVs. On the other hand, AV-1 was the hardest to mislead, with no more than 2.15\% of AEs by random AMG able to bypass its detection mechanisms.

\subsubsection{Combination of Generators}
In the following experiment, we created all conceivable pairs of all five generators, yielding 25 sets each of 2,000  adversarial malware examples. As described in Figure \ref{fig:method_diagram}, each set contains subsets $evasive_1$, $evasive_2$, and $failed_2$. The distribution of samples between the three subgroups depends on AV being tested. We submitted the generated AEs to all selected antivirus programs for further evaluation. 

\begin{figure}[h]
	\centering	
	\includegraphics[width=\linewidth]{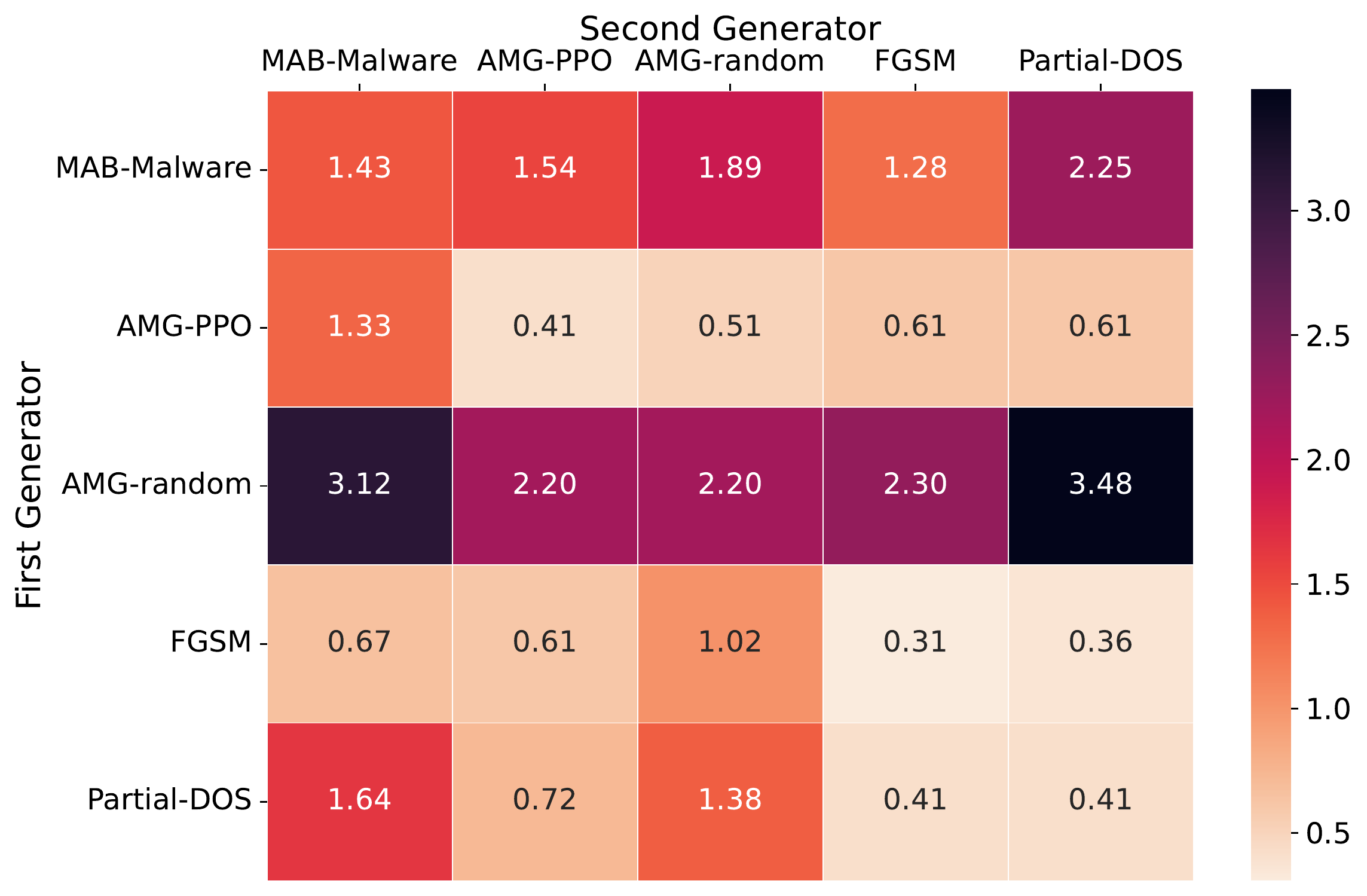} 
	\caption{Evasion rates of combined AE generators against AV-1.}
	\label{fig:AV-1_AE_generators_combinations}
\end{figure}

\begin{figure}[h]
	\centering	
	\includegraphics[width=\linewidth]{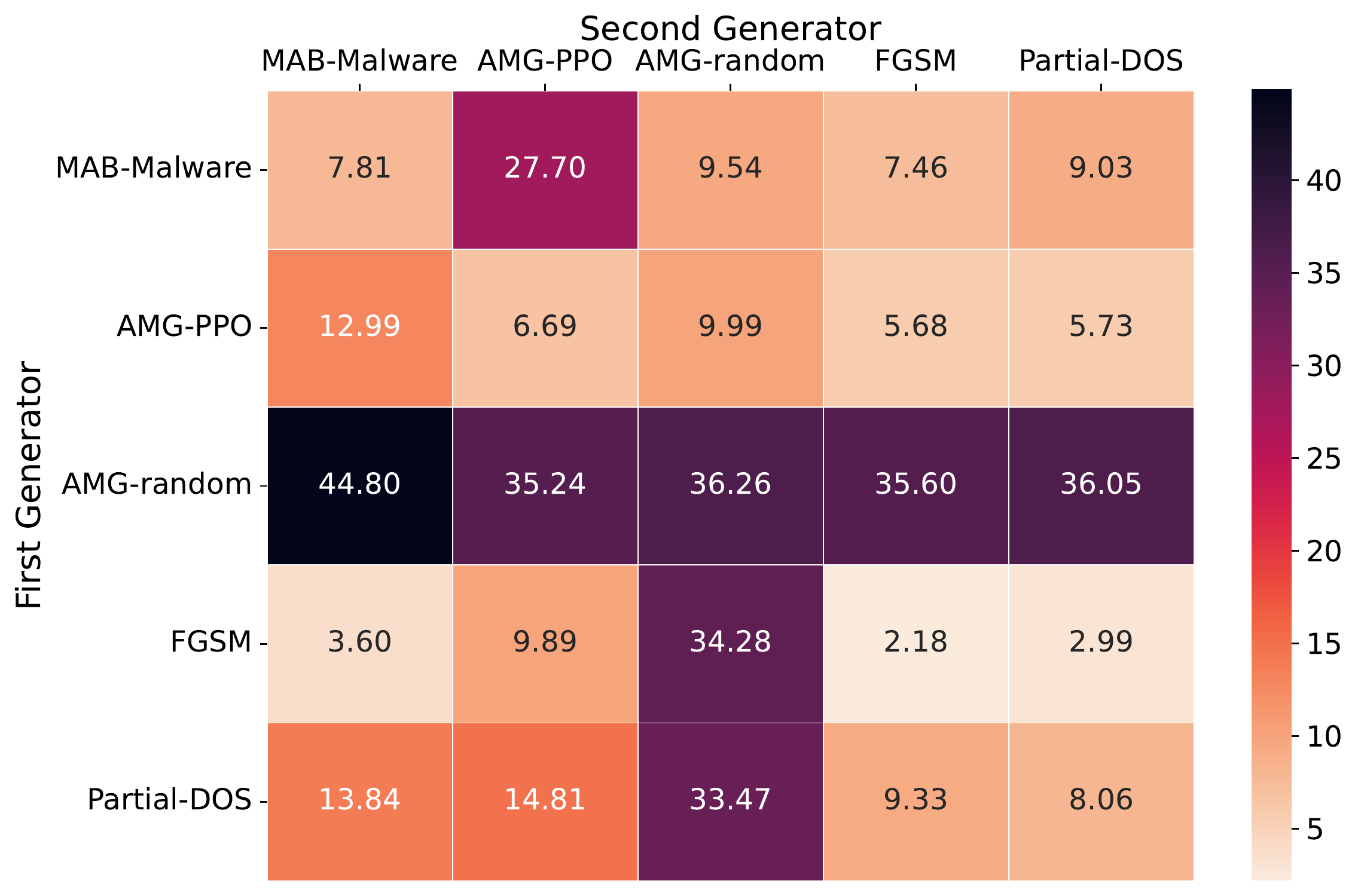} 
	\caption{Evasion rates of combined AE generators against AV-6.}
	\label{fig:AV-6_AE_generators_combinations}
\end{figure}


In Figures \ref{fig:AV-1_AE_generators_combinations} and \ref{fig:AV-6_AE_generators_combinations}, we can see the results of this testing for the AV-1 and AV-6 detectors, the detectors that performed best and worst in the baseline experiment, respectively. The first generator (vertical axis) was used for each of the 2,000 samples and produced AEs, from which the non-evasive AEs were used as input to the second generator (horizontal axis). The principal diagonal elements display repeated use of the same generator, i.e., the first and second generators are the same. If we compared the diagonal results with the previous testing listed in Table \ref{table:individual_AE_generators}, we can conclude that repeated use of the same generator leads to only marginal improvements. 

However, if the second generator differs from the first, we have seen significant improvements, as documented in Table \ref{table:combined_AE_generators_stats}. The relative minimal, maximal, and average values were calculated with respect to baseline values for each generator and AV product listed in Table \ref{table:individual_AE_generators}. The term ``relative'' refers to the percentage increase and was calculated as follows:

\begin{equation}
	relative = \dfrac{combined - baseline}{baseline} \cdot 100 \%
\end{equation}

\noindent where $combined$ denotes the evasion rate of combined generators and $baseline$ represents the evasion rate of a single generator. Note that the absolute and relative extremes were not necessarily recorded by the same combination of AE generators. 

\begin{table*}[h]
	\centering
	\caption{Absolute and relative evasion rate statistics recorded by combined AE generators when the first generator differs from the second one.}
	\label{table:combined_AE_generators_stats}
		\begin{tabular}{@{}l|ccc|ccc@{}}
\toprule
            & absolute min & absolute avg & absolute max & relative min & relative avg & relative max \\ \midrule
AV-1    & 0.31                 & 1.31                 & 3.48                 & 2.38                 & 102.75               & 356.91               \\
AV-2 	& 0.21                 & 5.56                 & 17.83                & 0.9                  & 487.72               & 5,600.0               \\
AV-3    & 0.7                  & 5.69                 & 15.46                & 0.0                  & 163.67               & 1,382.54              \\
AV-4    & 1.82                 & 8.45                 & 15.86                & 2.72                 & 477.17               & 3,516.67              \\
AV-5    & 0.26                 & 3.95                 & 13.59                & 0.0                  & 239.09               & 2,650.0               \\
AV-6    & \textbf{2.18}                 & \textbf{16.92}                & \textbf{44.8}                 & 2.21                 & 180.15               & 1,678.95              \\
AV-7    & 0.32                 & 3.76                 & 11.12                & 2.27                 & 238.07               & 1,125.0               \\
AV-8    & 1.9                  & 3.88                 & 11.24                & \textbf{2.9}                  & 71.78                & 492.5                \\
AV-9    & 0.16                 & 5.77                 & 17.95                & 0.0                  & \textbf{696.02}               & \textbf{8,650.0}               \\ \bottomrule
		\end{tabular}
\end{table*}

For the AV-1 detector, the highest absolute evasion rate of 3.48\% was recorded by the random AMG agent followed by the Partial DOS manipulations. The relative increase in evasion rates ranged from 2.38\% (AMG-random $\rightarrow$ AMG-PPO) to 356.91\% (Partial-DOS $\rightarrow$ MAB-Malware). Overall, AV-1 remained the most difficult antivirus to circumvent, with only 1.31\% of AEs becoming evasive on average. Similarly, the AV-6 antivirus continued to be the easiest to mislead, with up to 44.8\% of AEs generated by a combo of AMG-random followed by MAB-Malware being evasive. Across all combinations, an average evasion rate for the AV-6 detector was slightly below 17\%.

Even though our proposed attack struggled to generate successful AEs against the AV-1 detector, combining generators of adversarial malware examples, in this case, increased the success rate by more than 102\% on average compared to separate generators. The highest average evasion rate increase of almost 700\% was reported against AV-9, with a massive increase of 8,650\% recorded after extending the FGSM manipulations by the random AMG agent.

\begin{figure}[h!]
	\centering	
	\includegraphics[width=\linewidth]{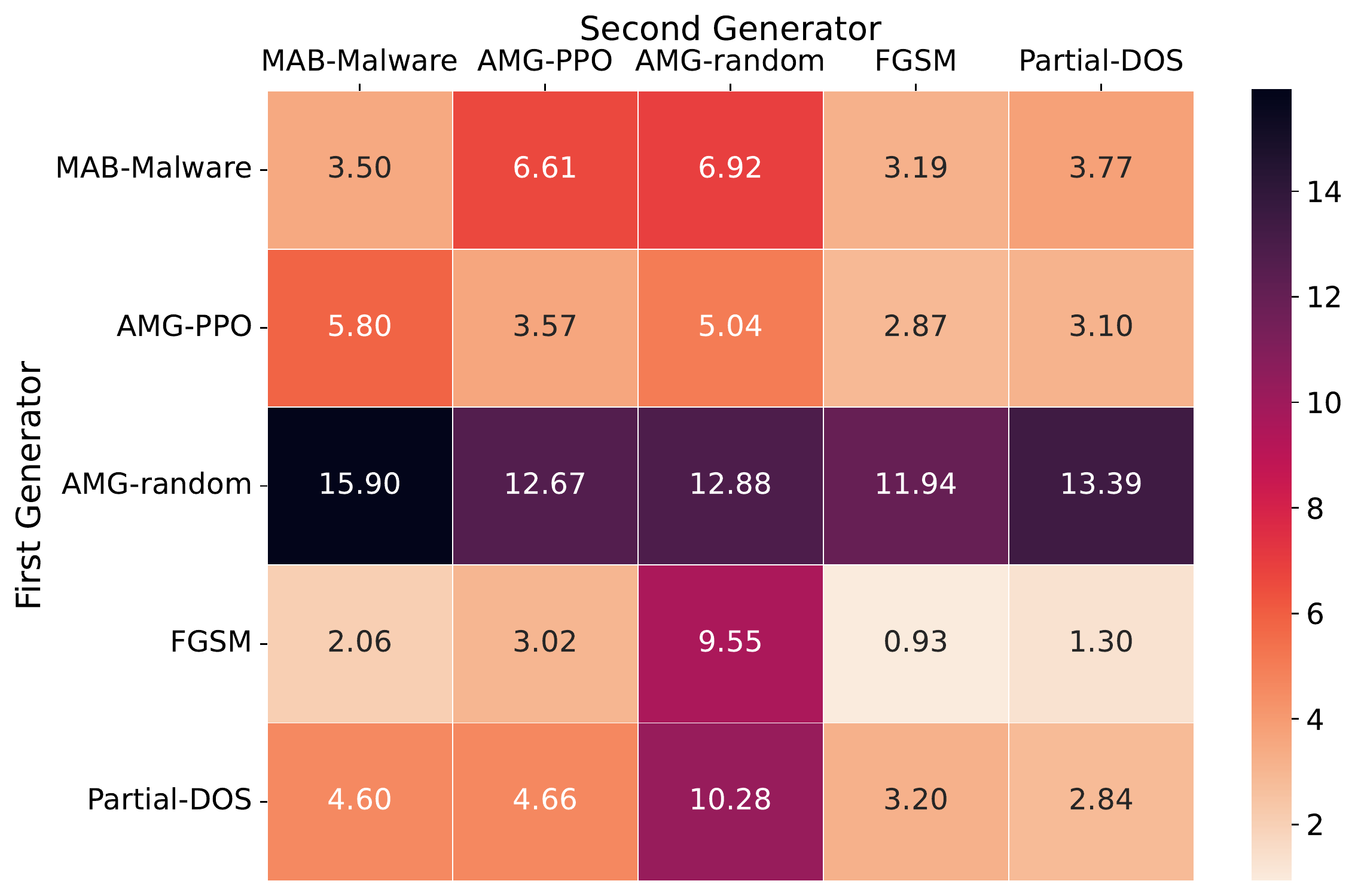} 
	\caption{Evasion rates of combined AE generators averaged over all tested AVs.}
	\label{fig:all_AE_generators_combinations}
\end{figure}

The average results against all tested AV programs can be found in Figure \ref{fig:all_AE_generators_combinations}. The results clearly show that the random AMG agent combined with any other generator outperformed almost all other combinations. Nevertheless, it is necessary to mention that some varieties of MAB-Malware and the trained AMG PPO agent achieved solid results as well. Overall, the most promising combination was the random AMG agent extended by the MAB-Malware generator, yielding an average evasion rate of 15.9\% against top AV products. When compared to the sole use of the AMG-random and MAB-Malware generators, this represents an average improvement of more than 36\% and 627\%, respectively.

\begin{figure}[h!]
	\centering	
	\includegraphics[width=\linewidth]{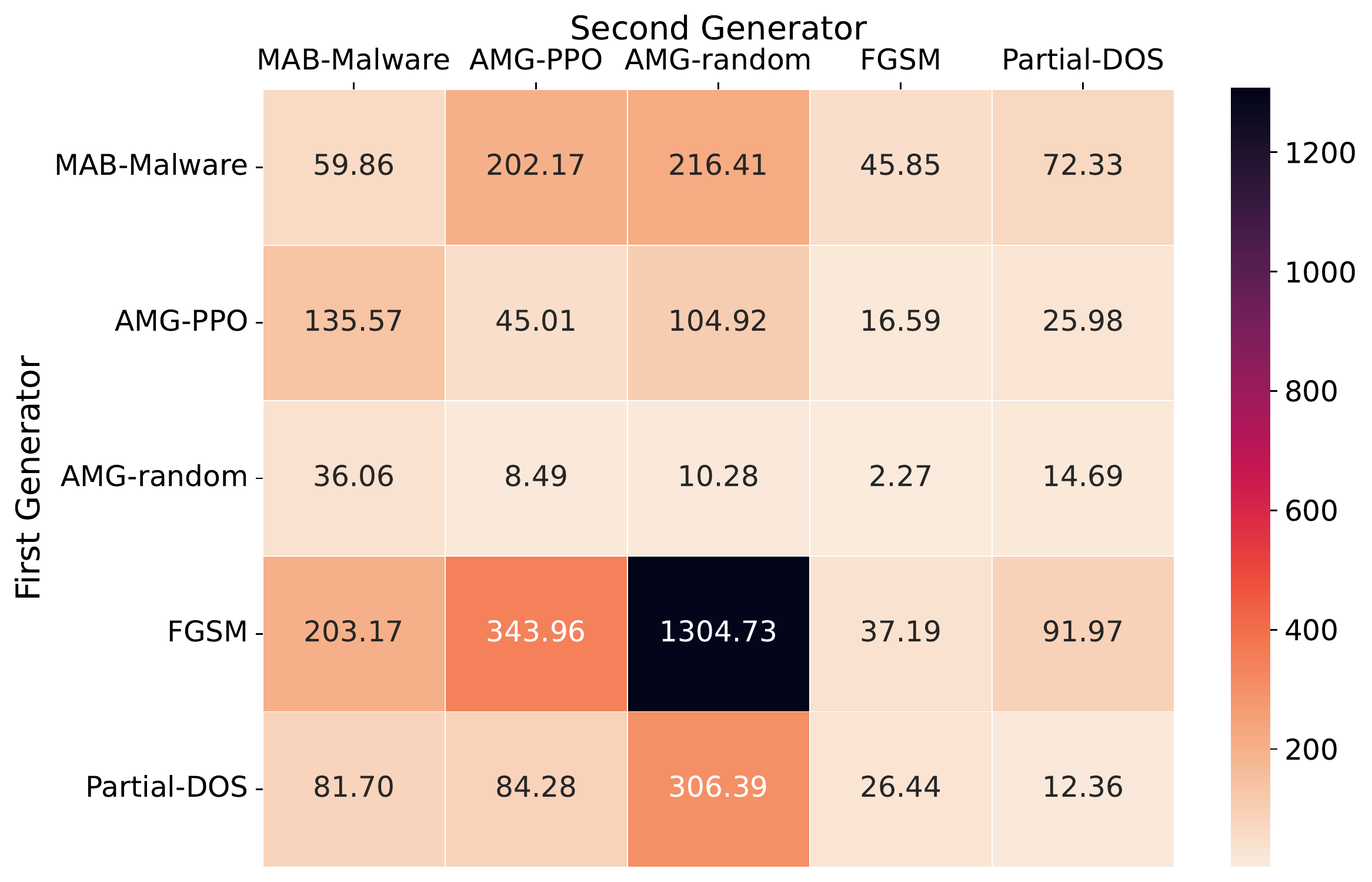} 
	\caption{The relative increase in evasion rates from a single AE generator to the combined pair averaged over all tested AVs.}
	\label{fig:relative_all_AE_generators_combinations}
\end{figure}

The average percentage improvements from using only the first generator to the combined pair against all AVs are shown in Figure \ref{fig:relative_all_AE_generators_combinations}. These results show that the FGSM injection attack benefits the most from extending its process of creating AEs by a different generator, e.g., by the random AMG agent, which led to an average improvement of over 1,304\%.

\section{\uppercase{Conclusions}} \label{sec:conclusion}
We proposed a novel method combining generators of adversarial malware examples to create more effective AEs. In total, we worked with five distinct AE generators, representing both black-box and white-box approaches. We evaluated this hypothesis on top-ranked antivirus programs and a set of 2,000 malware samples. Our proposed method is easy to implement and can significantly increase the evasion rate compared to those achieved by single generators.

Firstly, we measured the baseline readings of AEs generated by individual generators. The most successful AE generator was the random AMG agent with an evasion rate from 2.14\% up to 34.48\%, depending on the attacked antivirus. 

Next, we tested all possible pairs of individual generators on the same set of malware files and AV detectors. The results show that combining generators of adversarial malware increases the evasion rate significantly. For example, the random AMG agent extended by the MAB-Malware generator improved its evasion rate from 34.48\% against AV-6 to 44.8\% against the same antivirus. This combination, random AMG followed by MAB-Malware, proved to be the most successful among all tested combinations by achieving an average evasion rate of 15.9\% against leading AV products. On the other hand, the FGSM injection attack showed that substantial improvements could be made even for less-performing AE generators. Combined with different generators, this adversarial attack method saw massive improvements ranging from 91.97\% up to 1,304.73\%.

Based on our findings, we can conclude that the combination of AE generators improves the probability of bypassing anti-malware tools compared to single generators and shows that even top antivirus products are vulnerable to these attacks.

In the future, we would like to examine more in-depth if combining AE generators influences the resulting behavior of crafted AEs. In this work, we did not verify the functionality of resulting AEs but instead relied on the fact that the authors of AE generators create well-designed models. Thus connecting two AE generators should lead to functioning AEs. However, many authors of AE generators validate the preservation of original functionality only theoretically, but empirical results show that it is not sufficient \cite{kozak2023application}. 

Currently, our proposed method is only effective against static analysis detectors because it incorporates generators that only modify static analysis features. A challenging future research area would be to develop a reliable generator of AEs capable of bypassing dynamic analysis methods that could later be incorporated into our method.

\section*{\uppercase{Acknowledgements}}
This work was supported by the Grant Agency of the Czech Technical University in Prague, grant No. SGS23/211/OHK3/3T/18 funded by the MEYS of the Czech Republic and by the OP VVV MEYS funded project CZ.02.1.01/0.0/0.0/16 019/0000765 ``Research Center for Informatics''.

\bibliographystyle{apalike}
{\small
\bibliography{bib/mybibliography}}


\end{document}